\documentclass[useAMS]{mn2e}

\usepackage{amssymb}
\usepackage{amsmath}
\usepackage{tensor}
\usepackage{url}
\usepackage{hyperref}
\usepackage{wasysym}
\usepackage{subfigure} %permite usar subfigura 
\usepackage{graphicx}
\usepackage{epsfig} %permite recibir imagenes pdf
\usepackage{fancyhdr} %hace cabeceras bonitas
\usepackage{longtable}
\newcommand{\mincir}{\raise
-3.truept\hbox{\rlap{\hbox{$\sim$}}\raise4.truept\hbox{$<$}\ }}
\newcommand{\magcir}{\raise
-3.truept\hbox{\rlap{\hbox{$\sim$}}\raise4.truept\hbox{$>$}\ }}
\newcommand{\minmag}{\raise
-3.truept\hbox{\rlap{\hbox{$<$}}\raise5.truept\hbox{$<$}\ }}
\newcommand{\be}{\begin{equation}}
\newcommand{\ee}{\end{equation}}
 \newcommand{\ba}{\begin{eqnarray}}
\newcommand{\ea}{\end{eqnarray}}
\newcommand{\brr}{\begin{array}}
\newcommand{\err}{\end{array}}
\newcommand{\bc}{\begin{center}}
\newcommand{\ec}{\end{center}}

\newcommand{\hb}{\mathrm{H}\beta}

\newcommand{\Oiii}{\mathrm{O\ III}}

\newcommand{\logd}{\log_{10}}

\newcommand{\hii}{H\,\textsc{ii} }

\newcommand{\lsig}{$L(\mathrm{H}\beta) - \sigma$}

\title[HII galaxies and the Hubble constant]
{Determining the Hubble constant using Giant extragalactic HII regions and HII galaxies}
\author[R. Ch\'avez et al.]{Ricardo Ch\'avez$^{1}$, Elena Terlevich$^{1 * }$, Roberto Terlevich$^{1,2}$ 
\thanks{Visiting Professor UAM, Madrid} , Manolis Plionis$^{1, 3}$,
\newauthor Fabio Bresolin$^{4}$, Spyros Basilakos$^{5,6}$ and Jorge Melnick$^{7}$\\
$^1$ Instituto Nacional de Astrof\'\i sica, \'Optica y Electr\'onica,
Tonantzintla, Puebla, Mexico \\
$^2$ Institute of Astronomy, University of Cambrige, Cambridge, UK\\
$^3$ Institute of Astronomy \& Astrophysics, National Observatory of
Athens, Thessio 11810, Athens, Greece \\
$^4$  Institute for Astronomy, University of Hawaii, Honolulu, Hawaii, USA \\ 
$^5$ Academy of Athens Research Center for Astronomy \& Applied
Mathematics, Soranou Efessiou 4, 11-527 Athens, Greece \\
$^6$ High Energy Physics Group, Dept. ECM, Universitat de Barcelona,
Av. Diagonal 647, E-08028 Barcelona, Spain \\
$^7$ European Southern Observatory, Santiago de Chile, Chile 
}

\begin{document}

\maketitle

\begin{abstract}

We report the first results of a long term program aiming to provide
accurate independent estimates of the Hubble constant ($H_0$) 
%and the Dark Energy equation of state parameter ($w$) 
using the \lsig\ distance estimator for Giant  extragalactic \hii regions (GEHR) and \hii galaxies.

We have used VLT and Subaru high dispersion spectroscopic observations
of a local sample of \hii galaxies, identified in the
SDSS DR7 catalogue in order to re-define and improve the
$L(\mathrm{H}\beta)-\sigma$ distance
indicator and to determine the Hubble constant. To this
end we utilized as local calibration or `anchor' of this correlation, GEHR in nearby galaxies which have accurate distance
measurements determined
via primary indicators. Using our best sample of 69 nearby \hii galaxies  and 23 GEHR in 9 galaxies %with
                                %redshift independent distance
                                %estimation, %%MANOLIS CHANGE
we obtain $H_{0}=74.3 \pm$ 3.1 (statistical)$\pm$ 2.9 (systematic) km s$^{-1}$Mpc$^{-1}$, in excellent agreement with, and independently confirming,
the most recent SNe~Ia based results.
\end{abstract}


\vspace{1.0cm}

\section{Introduction}
The accurate determination of the Hubble constant, $H_0$, is considered one of
the most fundamental tasks in the interface between Astronomy and
Cosmology. The importance of measuring the expansion rate of the Universe to high
precision stems from the fact that $H_0$, besides providing cosmic distances, is
also a prerequisite %Its importance, beyond calibrating cosmic distances, 
%stems from the fact that the accurate determination of the local
%expansion rate of the Universe is a prerequisite 
for independent
constraints on the mass-energy content of the Universe  (e.g., Suyu et
al. 2012).
%, while its precise value is 
%used as a prior in cosmological parameter studies in order to break
%degeneracies.
%Indeed considering that the dark energy equation of state
%parameter evolves with time it has been found that $H_{0}$ lies in the 
%range of 61-84km s$^{-1}$Mpc$^{-1}$ at $2\sigma$ level (Tegmark et al. 2004).  
%The latter implies that there is 
%a strong degeneracy between $H_{0}$ and the other cosmological parameters 
%which means that the uncertainties on $H_{0}$ also reflect on 
%the values of $(\Omega_{m},w(z))$ and vice versa.

The direct determination of the
Hubble constant can only be obtained by measuring cosmic distances
%The only direct method to determine the Hubble constant is by measuring
%cosmic distances 
and mapping the local expansion of the Universe, since
the Hubble relation, $cz=H_0 d$, is valid and independent of the mass-energy content of the
Universe only locally ($z\mincir 0.15$). A variety of %such 
methods have
been used to estimate $H_0$, based on Cepheids,
surface brightness fluctuations, masers, the tip of the red giant branch (TRGB),
or type Ia supernovae [SNe~Ia] (for general reviews see Jackson 2007; Tammann, Sandage \&
Reindl 2008; Freedman \& Madore 2010). 
In particular, the use of SNe~Ia to measure the Hubble
constant has a long history in astronomy (eg., Sandage \& Tammann 1982; 1990). 
The subsequent discovery of the correlation between the magnitude at
peak brightness and the rate at which it declines thereafter (eg.,
Phillips 1993) allowed the reduction of the distance determination intrinsic
scatter.
However, one has to remember that SNe~Ia are secondary indicators and
their use relies on the determination of well-established local
calibrators, like the Large
Magellanic Cloud (LMC), Galactic Cepheids, the ``maser'' galaxy NGC 4258, etc. 
(cf. Riess et al. 2011).

Indirect methods to measure $H_0$ have also been developed (e.g. Bonamente et al. 2006; Suyu et al. 2010; Beutler et al. 2011), however, all of the indirect methods
use as priors other cosmological parameters, and thus the resulting
$H_0$ determinations are model dependent. 
% MANOLIS: WE CAN PUT THE FOLLOWING SENTENCE IF WE HAVE SPACE AT THE END
%For example, the WMAP7 analysis of 
%Komatsu et al. (2011) provide $H_{0}\simeq 70.4\pm 2.5$ 
%km s$^{-1}$ Mpc$^{-1}$, but in the context of a $\Lambda$CDM cosmological
%model. If however the priors of spatial flatness and a $w=-1$ 
%dark energy equation of state are relaxed,
%then the resulting $H_0$ span a wide range of values: $40\mincir
%H_0\mincir 80$ km s$^{-1}$ Mpc$^{-1}$ depending on the cosmological
%data used (eg. Riess et al. 2009a).
%%%%%%%%%%%%%%%%%%%%%%%%%%%%%%%%%%%%%%%%%%%%%%%%%%%%%%%%%%%%%%%%%%%%%%%%%

Returning to the direct method to estimate $H_0$, 
an important breakthrough occurred a decade or so ago by
the {\em HST} Calibration program (Saha et al. 2001; Sandage et
al. 2006) who found Cepheids in local galaxies that host SNe~Ia and
provided a Cepheid based zero-point calibration, and by the
{\em HST} Key project (Freedman et al. 2001) who furnished
a value of $H_{0}=72 \pm 2 {\rm (random)} \pm 7 {\rm (systematic)}$
km s$^{-1}$ Mpc$^{-1}$, based on Cepheid distances of external 
galaxies and the LMC as the first rung of the distance ladder.
This value was recently revised by the same authors, 
using a new Cepheid zero-point
(Benedict et al. 2007) and the new SNe~Ia of Hicken et al. (2009), to a
similar but less uncertain value of
$H_{0}=73 \pm 2 {\rm (random)} \pm 4 {\rm (systematic)}$
km s$^{-1}$ Mpc$^{-1}$ (see Freedman \& Madore 2010).
%Nevertheless, the above range of $H_0$ values was challenged by
Tammann et al. (2008) %who 
used a 
variety of local calibrators to recalibrate the SNe~Ia 
and found a significantly lower value of $H_{0}=62.3\pm 4$ km
s$^{-1}$Mpc$^{-1}$. The difference has since been explained as
being due to a variety of external causes among which the use of
heavily reddened Galactic Cepheids and of less accurate
photographic data (Riess et al. 2009a,b).

The most recent analysis of Riess et al. (2011) uses new {\em HST}
 optical and infrared 
observations of 600 Cepheid variables to determine the
distance to eight galaxies hosting recent SNe~Ia.
The resulting best estimate for the Hubble constant is:
$H_{0}=73.8\pm 2.4$ km s$^{-1}$ Mpc$^{-1}$ including random and
systematic errors.

From the above discussion it becomes clear that SNe~Ia 
are the only tracers of the 
%local 
Hubble expansion utilized to-date,
over a relatively wide redshift range ($0\mincir z \mincir 1.5$).
Therefore, due to the great importance of direct determinations of  the
Hubble constant for cosmological studies (eg., Suyu et al 2012) 
it is highly desirable to independently confirm the SNe~Ia based $H_0$ value by using
an alternative tracer.
%of the local Hubble flow. 

\hii galaxies have been proposed as such an alternative.
%$H_{0}=89\pm 10$ km s$^{-1}$ Mpc$^{-1}$. 
They  are massive and compact (in many cases unresolved) bursts of star formation in dwarf galaxies.
The luminosity of \hii galaxies is completely overpowered by that of the starburst. As a consequence they show the spectrum of a young \hii region, that indeed is what they are, hence their name.  Their similarity with GEHR is underlined by the fact that the first examples of prototype \hii galaxies, I~Zw18 and II~Zw40, were called ``Isolated Extragalactic \hii regions"  and found to be observationally indistinguishable from GEHR in nearby galaxies (Sargent and Searle 1970). They are discovered mainly in spectroscopic surveys due to their strong narrow emission lines, i.e. very large equivalent widths.

It is important to emphasise that the optical properties of \hii galaxies are those of the young burst with almost no information (or contamination) from the parent galaxy. This is a direct consequence of selecting \hii galaxies as those systems with the largest equivalent width (W) in their emission lines, i.e. W(H$\beta) > $ 50\AA .

Because the starburst component can reach very high luminosity,  \hii galaxies  can be observed at large redshifts ($z > 3$).
What makes these galaxies interesting cosmological distance
probes (cf. Melnick, Terlevich \& Terlevich 2000 ; Siegel
et al. 2005) is the fact that as the mass of the starburst component
increases, both the number of ionizing photons and the turbulent
velocity of the gas, which is dominated by the star and gas  gravitational
potential, also increases. This induces a correlation between the
luminosity of recombination lines, e.g. $L(\mathrm{H}\beta)$
and the ionized gas velocity dispersion $\sigma$ (see
Terlevich \& Melnick 1981; Hippelein 1986; Melnick, Terlevich \& Moles 1988; Melnick, Terlevich \& Terlevich 2000;
Fuentes-Masip et al. 2000; Telles et al. 2001, Bosch et al. 2002;
Siegel et al. 2005; Bordalo \& Telles 2011).

 A first attempt to estimate $H_0$, using \hii galaxies and GEHR as local
calibrators, was presented in Melnick, Terlevich \& Moles (1988).
The use of \hii galaxies as deep cosmological tracers was discussed by Melnick, Terlevich \& Terlevich (2000) and Siegel et al. (2005). 
Recently, we presented a thorough investigation
of the viability of using \hii galaxies to constrain the dark energy
equation of state, accounting also for the effects of gravitational lensing, which
are expected to be non-negligible for very high redshift `standard candles' and
we showed that indeed \hii galaxies can provide an important cosmological probe
(Plionis et al. 2011).%accounting also for the effects of gravitational
%lensing, expected to be considerable for very high redshift 'standard
%candles' and they indeed appear to be a prominent cosmological probe
%(Plionis et al. 2011).

The aim of the current paper is to use \hii galaxies and a local
calibration of the \lsig\ relation based on GEHR of nearby galaxies, as an alternative
direct approach for estimating the Hubble constant over a redshift
range of $0.01<z< 0.16$.

\section{Sample selection and observations}

A sample of 128 \hii galaxies was selected
from the SDSS DR7 spectroscopic data release (Abazajian, et al. 2009) within 
a redshift range $0.01<z<0.16$, chosen for being compact ($D<5\ \mathrm{arcsec}$), having 
large Balmer  emission line fluxes and  equivalent widths. %An equivalent width lower limit of 50 \AA\ for  $\hb$  
A lower limit for the equivalent width of $\hb$  of 50 \AA\ was chosen  to avoid starbursts that are either evolved
or  contaminated by an underlying  older
stellar population component (cf. Melnick, Terlevich \& Terlevich 2000).
The redshift lower limit was chosen to minimize the effects of local
peculiar motions relative to the Hubble flow and the upper limit to 
minimize any possible Malmquist bias and to avoid gross 
cosmological effects. %The $z$ distribution of the sample is shown in figure 1.

%Large $W$ were chosen 
%in order to avoid more evolved starbursts, that
%would present underlying absorptions due to an older stellar
%population component, thus affecting the emission lines flux
%[cf. Melnick, Terlevich \& Terlevich (2000)].

In order to improve the parameters of the \lsig\  relation obtained 
from previous work, high-resolution echelle spectroscopy for the \hii galaxy sample was
performed at 8 meter class telescopes. We used  the Ultraviolet and
Visual Echelle Spectrograph (UVES)  (Dekker et al. 2000) at the
European Southern Observatory (ESO) Very Large Telescope (VLT) in
Chile, and the High Dispersion Spectrograph (HDS) (Noguchi et
al. 2002; Sato et al. 2002) at the National Astronomical
Observatory of Japan (NAOJ) Subaru Telescope on Mauna Kea, Hawaii. The chosen setups
provided UVES spectra centred at $5800\ \mathrm{\AA}$ with a
slit-width of 2\arcsec, giving a spectral resolution of $\sim 22000$. The HDS spectra were centred at $\sim 5400\
\mathrm{\AA}$, and with a slit width of  4\arcsec\ the spectral
resolution obtained was  $\sim 9000$.

To obtain accurate total $\hb$ fluxes for the \hii galaxy
sample, we performed long slit spectrophotometry  at 2-meter class
telescopes  under photometric conditions and using a slit width  (8
arcsec) larger  than the upper limit of the HII galaxies size in our
sample. We used the Boller \& Chivens spectrographs
at the 2.1 m telescope of the 
Observatorio Astron\'omico Nacional (OAN) in San Pedro
M{\'a}rtir and at the 2.1 m telescope of the
Observatorio Astrof\'isico Guillermo Haro (OAGH) in Cananea, both in M\'exico.

Full details of the sample selection, observations and data reduction
and analysis are given elsewhere (Ch\'avez et al., in
preparation). Here we summarize the relevant results regarding the
determination of the distance estimator and $H_{0}$.

$\hb$ and  $[\Oiii]\ \lambda \lambda 4959,\ 5007$ line widths were
measured fitting single gaussians to the line profiles. 
As previously found most  \hii galaxies show line profiles that are well fitted by single gaussian (e.g. Melnick et al. 1988, Bordalo \& Telles 2011).
We cleaned the sample by first removing from the original list  those \hii galaxies  with either asymmetric or double/multiple line profile. We also  removed those  \hii galaxies showing rotation or large photometric errors in their $\hb$ fluxes or with an uncertain reddening correction. 
All this reduced the sample from 128 to 69 \hii galaxies.

%Figure~4? shows some examples of the line profiles of rejected objects. 

The values of the observed velocity dispersions, $\sigma_o$, were
corrected for thermal ($\sigma_{t}$) and instrumental ($\sigma_{i}$)
broadening, and the final corrected dispersion was estimated according
to:
\begin{equation}\label{eq:sig}
 \sigma = (\sigma_o^2 - \sigma_t^2 - \sigma_i^2)^{1/2}\;.
\end{equation}

The $1\sigma$ uncertainties of the velocity dispersion were estimated from multiple observations computing  the variance over
the repeated measurements; otherwise as the mean value of the obtained relative errors.

$\hb$ integrated fluxes were measured by fitting a single gaussian to
the long slit spectra, while their $1\sigma$ uncertainties were estimated from
the expression (e.g. Tresse et al. 1999):
\begin{equation}
 \sigma_{F} = \sigma_{c} D (2 N_{pix} + W/D)^{1/2}\;,
\end{equation}
where $\sigma_{c}$ is the mean standard deviation per pixel of the
continuum on each side of the line, $D$ is the spectral dispersion,
$N_{pix}$  is the number of pixels covered by the line and $W$ is the
line equivalent width.

%RICARDO. It is necessary to measure the line fluxes by integration of the line profile. It should be a better measure than the gaussian fit 

Heliocentric redshifts and their uncertainties were obtained from the
SDSS DR7 and DR8 spectroscopic data when available, otherwise from our echelle data or the Spectrophotometric Catalog of \hii galaxies 
(Terlevich et al.~1991). The redshifts have been transformed from the
heliocentric to the local group reference frames following Courteau \& van den
Bergh (1999) and corrected for the local bulk flow
using the model of Basilakos and Plionis (1998).  The $1\sigma$
uncertainties were propagated using a Montecarlo procedure.

%(details will appear elsewhere; Ch\'avez et al. in preparation).

To determine the zero point  for the \lsig\ relation, we obtained data
from the literature for a sample of 23 GEHR in 9 nearby galaxies whose distances
have been measured by means of well tested primary distance
indicators.

%namely Cepheids,  RR Lyrae and Eclipsing Binaries. 

The details of the GEHR data will also be given in Ch\'avez et al. (in
preparation). For these objects, velocity dispersions have been taken
from Melnick et al. (1987), whereas distance moduli have been
obtained averaging over the available measurements published after 1995,
selecting only those based on  Cepheids, RR Lyrae, Mira variables and eclipsing
binaries except for those in IC~2574 and NGC~4236 for which
only TRGB 
measurements are available. The adopted distance moduli ($ \mu$) are listed as an inset in Figure \ref{fig:LSgeh2r}.
The global integrated $\hb$ 
fluxes and corresponding extinction were obtained 
%as weighted averages (based on observations accuracy) of 
from the values reported by  Melnick et al. (1987). 

\section{Determination of $H_{0}$}
The procedure we use to estimate the Hubble constant comprises  three steps:

%%%%%%%%%%%%%%%%%% Fig. 1 %%%%%%%%%%%%%%%%%%%
%Figure --- L - Sigma Relation for the GEH2R sample
\begin{figure}
\centering
\resizebox{8cm}{!}{\includegraphics{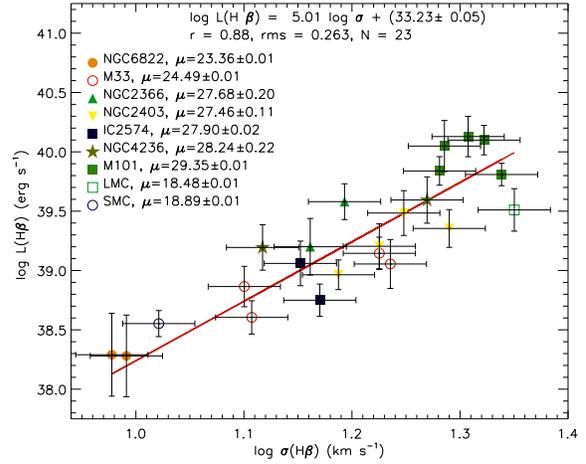}}
\caption {\lsig\ relation for the GEHR sample. The correlation parameters and the adopted individual distance moduli are given in the inset. 
The line is the best fit for   the slope determined by the fit to the HII galaxies (see Eq. 3).}
\label{fig:LSgeh2r}
\end{figure}
%%%%%%%%%%%%%%%%% Fig. 2 %%%%%%%%%%%%%%%%%%%%
%%%%%%%%%%%%%%%%%%%%%%%%%%%%%%%%%%%%%
%Figure --- L - Sigma Relation for the GEH2R+HII galaxies sample
\begin{figure}
\centering
\resizebox{8cm}{!}{\includegraphics{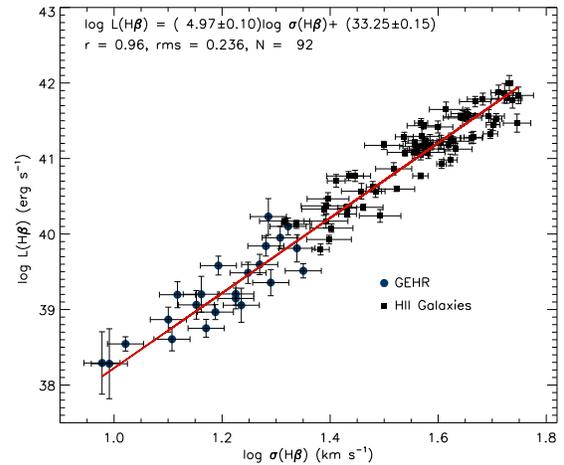}}
\caption {\lsig\ relation for the joint \hii galaxies and GEHR samples. The least square fit considers the errors in both axes.}
\label{fig:LSigJoin}
\end{figure}
%%%%%%%% %%% Important
%%%%%%%%%Now working in a new version of this plot
%%%%%%%%%%%%%%%%%%%%%%%%%%%%%%%%%%%%%

\begin{enumerate}
\item First we determine the slope of the
$L(\mathrm{H}\beta)-\sigma$ relation for \hii galaxies.
Since the slope is independent of $H_0$ we use an arbitary value of $H_0$ to determine luminosities
from the observed $\hb$ flux and the Hubble distance\footnote{We have
  verified that the initial choice for the value of $H_0$ does not alter the
  determined slope value.}.

\item  We then
%Once we have determined the value of the slope, the second step
%consists in determining the
determine the intercept of the relation from a fit to the `anchor'
GEHR sample, but fixing the slope to that determined in step one, i.e., that based on \hii galaxies.
Figure \ref{fig:LSgeh2r} 
shows the \lsig\ relation for the GEHR
sample. The slope of the correlation has been fixed to the value obtained 
from the \hii galaxies sample fitting in (i).
%The thick blue points show the weighted -- based on the
%measurement accuracy -- mean values for each
%galaxy, whereas the thin points are the values for individual GEHR.

The  resulting  \lsig\ correlation for the joint sample of GEHR and \hii galaxies is:
\begin{equation}
 \logd L(\mathrm{H}\beta) = (4.97 \pm 0.10)   \logd \sigma + (33.25 \pm 0.15)
\end{equation}
has r.m.s.~$ \logd L(\mathrm{H}\beta)= 0.236$ and is shown in figure \ref{fig:LSigJoin} .

%The \lsig\ correlation resulting from fitting jointly the \hii galaxies sample and the
%GEHR sample is shown in Figure \ref{fig:LSigJoin} and is given by:
%The resulting \lsig\ correlation, calculated as above described and
%shown together with the GEHR and \hii galaxy samples in 
%Figure 3, is given by:
%\begin{equation}
 %\logd L(\mathrm{H}\beta) = (4.97 \pm 0.10) \logd \sigma + (33.26 \pm 0.15) 
%\end{equation}
%with r.m.s. $\logd L(\mathrm{H}\beta)= 0.234$.
%\ref{fig:LSjoint}. 
%The parameters of the fit are given as
%labels in both figures, and the procedure is  explained in the
%following section.

\item Finally we determine the value of $H_0$
by minimizing, over a grid of $H_0$ values, the function:
\begin{equation}
\chi^2(H_0) = \sum\limits_{i=1}^n \frac{[L_i(\sigma_i) -
  \tilde{L}_i(H_0, f_i, z_i)]^2}{\Delta_{L, i}^2 + \Delta_{\tilde{L}, i}^2}, 
\end{equation}

\noindent where the summation is over the \hii galaxies, 
$\sigma_i$ are the measured velocity dispersions (eq. \ref{eq:sig}),
$L_i(\sigma)$ are the luminosities estimated from the `distance
indicator' as defined in equation~3,  $\Delta_{L, i}$ are their errors
propagated from the uncertainties in $\sigma$ and the slope and intercept
of the relation. $\tilde{L}_i(H_o, f_i, z_i)$ are the luminosities obtained from
the measured fluxes and redshifts  by using a particular value of
$H_0$ in the Hubble law to estimate distances, and
$\Delta_{\tilde{L}, i}$ are the errors in this last estimation of
luminosities, propagated from the uncertainties in the fluxes and
redshifts.   

\end{enumerate}
%%%%%%%%%%%%%%%% Fig. 3 %%%%%%%%%%%%%%%%%%%%%
%Figure --- Chisq plot
\begin{figure}
\centering
\resizebox{8cm}{!}{\includegraphics{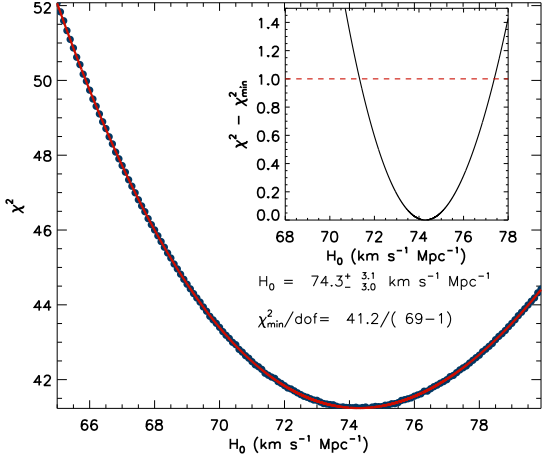}}
\caption {Values of  $\chi^2$  for the grid of $H_0$. The solid line is a cubic fit to the points. 
The inset panel shows the value of $\chi^2 - \chi^2_{min}$.}
\label{fig:chsqplot}
\end{figure}
%%%%%%%%%%%%%%%%%%%%%%%%%%%%%%%%%%%%%
Figure \ref{fig:chsqplot} 
shows the resulting $\chi^2$ for the range of
$H_0$ values used, with the solid line being a cubic fit to the points. 
The $1\sigma$ confidence limits of  $H_0$  were obtained from the values for which  $\chi^2 - \chi^2_{min} = 1$ since the fit has only one degree of freedom (see the  inset panel in Figure \ref{fig:chsqplot}).

The value obtained for $H_0$ using the above described procedure is:
\begin{equation}
  H_0 = 74.3^{+3.1}_{-3.0}\ \mathrm{km\ s^{-1}\ Mpc^{-1}}\;.
\end{equation}

\begin{figure}
\resizebox{8.75cm}{!}{\includegraphics{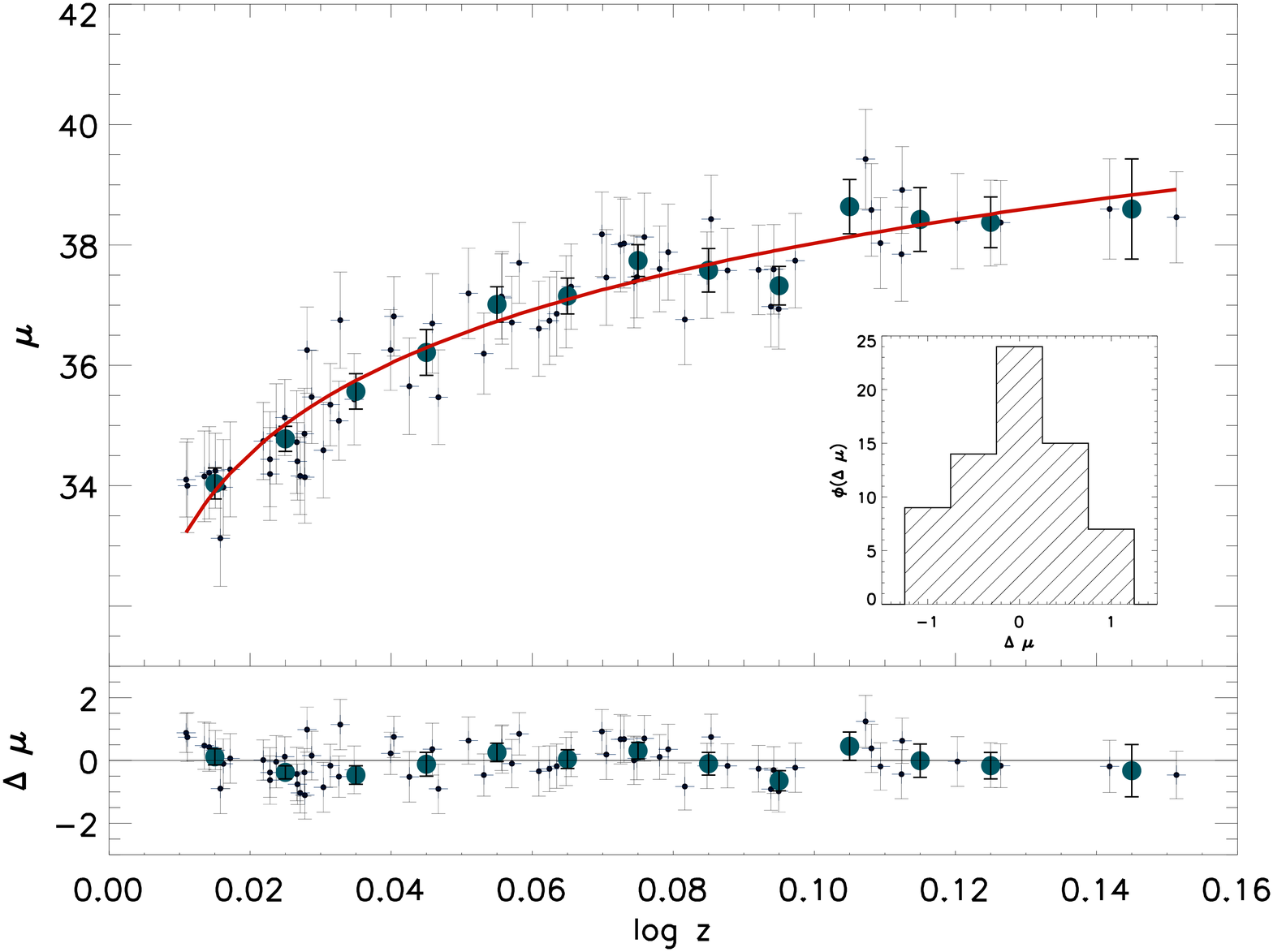}}
\caption {Hubble diagram for our sample of  69 \hii galaxies. The thick points
are the mean values for bins of $0.01$ in redshift. The solid line shows the run with redshift of the distance modulus for $H_{0}$= 74.3. Residuals are plotted in the bottom panel and their distribution is shown in the inset. The r.m.s. value is 
0.57 mag.}
\label{fig:HRsh}
\end{figure}

Figure \ref{fig:HRsh} shows the Hubble diagram
for the sample of \hii galaxies used for the $H_0$ value
determination. The continuous line
 shows the redshift run of the distance modulus, obtained from the linear
 Hubble law and the fitted $H_0$ value, whereas the points correspond
 to the individual \hii galaxy distance moduli %from the luminosities 
obtained  through the \lsig\ correlation.

The quoted Hubble constant uncertainty in equation 5 reflects only the random
errors, while   systematic errors can also affect the mean value as well as
the overall $H_0$ uncertainty. We have identified as potential sources
of systematic errors the following:
%\begin{enumerate}
(a) the broadening of the emission lines, being contaminated by a rotational velocity component, 
(b) the internal structure/multiplicity  of GEHR and \hii galaxies, 
(c) stellar winds affecting the line profiles
(d) internal extinction, 
(e) coherent or peculiar motions affecting the redshifts of nearby \hii galaxies, 
(f) the age  of GEHR and \hii galaxies, 
(g) the Malmquist bias, 
(h) variations in the IMF. 
%\end{enumerate}
The detailed discussion of these systematics is an important aspect of the $H_0$ determination and will be presented 
in a future paper (Ch\'avez et al.  in preparation). Here we briefly discuss these systematics, and the procedures used to minimize them.

(a) and (b)  To minimize the rotation and multiple component systematic
effect we have used in our correlation only those objects with
emission line profiles that are Gaussian and show no multiple
components [for a discussion see Bosch et al. (2002)].

(c) The presence of weak extended (non-gaussian) wings in the emission
line profiles introduces a small systematic effect. These weak wings
are probably associated with stellar winds. The resulting effect is
that taking into account the wings in the fit the final FWHM tends to
be slightly smaller. This should affect similarly both GEHR and \hii\
galaxies. We estimate that this may introduce a systematic error of about 2
percent in $H_0$.

(d) The extinction has been always estimated using the Balmer
decrement method. We do not expect a sizeable systematic effect associated
with this correction.

(e) \hii galaxies tend to populate the voids so local
peculiar motions should be relatively small. Furthermore, to
  minimize the effects of coherent bulk flows on the redshifts of
  the \hii galaxies, we imposed a lower radial velocity limit of
  3000 km/s. In any case, we have also computed $H_{0}$
  including a local bulk flow correction and found no overall effect.
  
%%%%%%%%%%%%% Fig. 4 %%%%%%%%%%%%%%%%%%%%%
%Figure --- Hubble Diagram
%%%%%%%%%%%%%%%%%%%%%%%%%%%%%%%%%%%%

 (f) The age of the GEHR and \hii galaxies affects their M/L ratio and
 therefore the zero point of the \lsig\ relationship. To minimize this
 effect we have specifically selected objects with  $W(\hb) > 50\
 \mathrm{\AA}$. This guarantees that the age of the star forming
 region is less than 6 Myr, thus minimizing the effect of evolution 
 (Leitherer et al. 1999). We estimate that at most a plausible
 systematic difference in ages between GEHR and \hii galaxies may
 affect $H_0$ at a 2\% level.

 (g) We have %corrected  $H_0$
 calculated the Malmquist bias following the procedure proposed by
 Giraud (1987) adopting a power law luminosity function, 
with a slope $\alpha = -1.7$. We
 have obtained a value of $2.1\ \mathrm{km\ s^{-1}\ Mpc^{-1}}$ at 
$z=0.16$, which we consider as one of the systematic error components.
 % After applying the correction for the $H_0$ value calculated up to $z = 0.1$ we obtain:
% \begin{equation}
%  H_0 =  73.6^{+3.0}_{-2.9} \pm 1.9 \ \mathrm{Km\ s^{-1}\ Mpc^{-1}}\;,
% \end{equation}
% where we have corrected the central value for the Malmquist bias
% effect and added as a systematic error the magnitude of the
% correction. 

 (h) The \lsig\ distance estimator relies on the universality of the
 IMF. Any systematic variation in the IMF will affect directly the M/L ratio
 and therefore the slope and zero point of the relation. The fact that
 our estimates of the Hubble constant are in agreement with those from
 SN~Ia supports the hypothesis of a universal IMF.

%%%%%%%%%%%%%%%%%%%%%%%%%%%%%%%%%%%%%
 %Table --- Systematic error budget 
\begin{table}{

% \centering
 %\begin{minipage}{3.0in}
  \caption {Systematic error budget on the $H_0$ determination }
 %\resizebox{0.4 \textwidth}{
\begin{tabular}[h] { l l l }
\hline
\hline
Symbol & Source & Error (km s$^{-1}$ Mpc$^{-1}$)  \\
\hline
$\sigma_a,b$ & Rotation,Multiplicity & 0.7 \\
$\sigma_c$ & Stellar Winds & 1.1\\
$\sigma_d$ & Internal Extinction & 0.7 \\
$\sigma_f $ & Object's Age & 1.4 \\
$\sigma_g$ & Malmquist Bias & 2.1 \\
$\sigma_h$ & IMF & ---- \\
%\hline 
%Subtotal, $\sigma_{Ho}$ & & 4\\
%\hline
%Analysis Systematics & & ?\\
\hline
%Total sys. $\sigma_{H0}$  &  &3.4\\
Total &  &2.9\\

\hline

\label{tab:SEB}
\end{tabular}
}
%\end{minipage}

\end{table}
%%%%%%%%%%%%%%%%%%%%%%%%%%%%%%%%%%%%%

Table \ref{tab:SEB} shows the systematic error budget on the $H_0$
determination.

\section{Conclusions}
It is indisputable that in the epoch of intense studies aimed at measuring the
dark energy equation of state, it is of paramount importance to
minimize the amount of priors needed to successfully complete such a
task. One such prior is the Hubble constant $H_0$  and  its
measurement at the $\sim 1\%$ accuracy level  has been identified as a
necessary prerequisite for putting effective constraints on the dark
energy, on neutrino physics and even on tests, at cosmological scales,
of general relativity (see Suyu et al. 2012). Furthermore, it is highly desirable to have independent
determinations of $H_0$, since this will help understand and control
systematic effects that may affect individual methods and tracers of
the Hubble expansion.

It is within this latter strategy that our current work falls. We have carried out
%It is within this latter strategy that our current work falls, for
%which we have performed 
VLT and Subaru observations of a sample of nearby
\hii galaxies identified in the
SDSS DR7 catalogue and 2m class telescopes spectrophotometry, in order to define their $L(\mathrm{H}\beta)-\sigma$ 
correlation, which we use to estimate the value of the Hubble
constant. This is achieved by determining the zero-point of the distance
indicator using GEHR in
nearby galaxies, for which accurate independent distance measurements
exist (based on Cepheids, RR Lyrae, TRGB  and eclipsing binaries). 

Using our sample of  92 objects (69  \hii galaxies with 
$z\mincir 0.16$ and 23 GEHR in 9 galaxies with distances determined via primary indicators) we obtain: \\ $H_{0}=74.3\pm 3.1$(random) $\pm 2.9$ (systematic)  km s$^{-1}$
Mpc$^{-1}$,\\ 
in excellent agreement with, and independently confirming,  the recent SNe~Ia-based
results of Riess et al. (2011).

\section*{Acknowledgements} 
The authors thank the support by VLT, Subaru, San Pedro
M\'artir and Cananea Observatories staff  and the hospitality of the
Departamento de F\'\i sica Te\'orica of the Universidad 
Aut\'onoma de Madrid, where part of this work was done.
RC, RT, ET and MP acknowledge the Mexican Research Council CONACYT for
financial support through grants 
CB-2005-01-49847, CB-2007-01-84746 and CB-2008-103365-F.
FB acknowledges partial support from the National Science Foundation
grants AST-0707911 and AST-1008798.
S.B.
% wishes to thank the Dept. ECM of the
%University of Barcelona for hospitality, 
and RT acknowledge
financial support from the
Spanish Ministry of Education, within the program of Estancias de
Profesores e Investigadores Extranjeros en Centros Espa\~noles (SAB2010-0118 and SAB2010-0103).
We thank the anonymous referee whose  suggestions greatly improved the clarity of this letter.

\end{document}